\documentclass[12pt]{article}
\usepackage{amssymb}

\def\beq{\begin{equation}}
\def\eeq{\end{equation}}

\newcommand{\ds}{\displaystyle}
\newcommand{\BI}{Born-Infeld}
\newcommand{\st}{\!*\!}

\begin{document}

\begin{flushright}

ITEP/TH-21/02\\

\end{flushright}

\vspace{0.5cm}

\begin{center}

{\Large\bf
Helicity conservation in Born-Infeld theory
\footnote{Talk given at the Workshop "String Theory and Complex Geometry"
(Bad Honnef, 8th-12th April, 2002)}}\\

\end{center}

\bigskip

\begin{center}

{\bf A.A.Rosly and K.G.Selivanov}

\bigskip

{ ITEP, Moscow, 117218, B.Cheryomushkinskaya 25}

\end{center}

\bigskip

\begin{abstract}
We prove that the helicity is preserved in the scattering of photons
in the Born-Infeld theory (in 4d) on the tree level.
\end{abstract}

1. The Born-Infeld theory \cite{born} introduced at the outset of the field
theory epoch
as a non-linear
generalization of Maxwell theory has shown up recently as
an effective theory of D-branes \cite{tseitlin}.
From the perturbative point of view, the Born-Infeld theory is quite
complicated
since its Lagrangian is an infinite power series in
$F^2=F^{\mu \nu}F_{\mu \nu}\,$, so that
there are infinitely many vertices,
which make almost hopeless any attempt to directly analyse the corresponding
Feynamn diagrams,
to say nothing on the issue of renormalizability.
Nevertheless, it might be interesting to notice that the \BI\ theory in four
dimensions possesses certain curious properties. Recall that the 4d Maxwell
equations, $dF=d\st F=0$, are invariant with respect to the duality
transformations: $\delta F=*F$. These transformations extend to the non-linear
\BI\ theory \cite{duality}.

Below, we make use of these transformations to deduce the conservation
of helicity in the tree amplitudes of the \BI\ theory.  It might be
also worthy pointing out that such a conservation law does not appear
to be related with any symmetry of the theory: the duality
transformations are defined on-shell only and do not correspond to a
symmetry of the Lagrangian. Therefore there are also no immediate
consequences for the loop amplitudes (cf., however, a discussion in
the conclusion). The remarkable helicity conservation (or, better to
say, the selection rule) could indicate that the \BI\ theory is
solvable in a sense. As a matter of fact, it is not only possible to
prove the helicity conservation in this theory, but also give a closed
analytic expression for all the tree amplitudes (with any number of
particles) as we shall show elsewhere \cite{RS2}.

\medskip

2. The \BI\ theory is the theory of the abelian gauge potential $A$
with the Lagrangian
\beq\label{lagr}
L=\sqrt{det(g+F)}\,,
\eeq
where $F=dA$ is the field strength 2-form and $g$ is the (flat) space-time
metric\footnote{
We shall consider this field theory in flat space-time only and adopt
the convention $*^2=1$ for the duality operator $*$, disregarding the
conventions related to the signature of the metric and of the reality
of the fields: for our aims we shall anyway need complex fields.}.
The corresponding field equations along with Bianchi identities read as
\beq\label{eqm}
dF=0\;,~~ d\st D=0\,,
\eeq
where $D=\partial L/\partial F$. Let us consider the following infinitesimal
transformation:
\beq \label{it}
\delta F = *D \,.
\eeq
Note that $D$ is by definition a certain function of $F$, namely,
$D(F)=\partial L/\partial F$, and therefore, the transformation law of $D$
follows from that of $F$. In fact, one can verify \cite{duality} that
\beq
\delta D = *F \,.
\eeq
It is now obvious that the field equations (\ref{eqm}) are invariant with
respect to such duality rotations. Note that the transformation
$\delta F = *D$ is a non-local non-linear transformation of the fundamental
field $A$ of the theory and is defined only ``on the mass shell'':
since $F=dA$ implies $dF=0$ as an identity, $\delta F = *D$ is defined
in terms of $A$ if only $d\st D=0$, i.e.\ on-shell.

\medskip

3. In scattering theory we have to determine the quadratic (free-field)
part of the Lagrangian and of the field equations. In our case, the latter
are, of course, just the Maxwell equations for the gauge potential $a$,
\beq
df=0 \;,~~ d\st f=0 \,,
\eeq
where $f=da$. The plane wave solutions to these linear equations,
\beq\label{pw}
a=h\,e^{ik\cdot x} \,,
\eeq
are defined by a 4-momentum $k$ and a polarization 4-vector $h$, where
$k^2=0,~ k\cdot x=k_\mu x^\mu\,$, and $k\cdot h=0$ (the Lorentz gauge).
It is convenient to use the basis of self-dual and anti-self-dual plane waves,
corresponding to $f=\pm*f$ and assign a quantum number, the {\it self-duality
number},
$s=1$ to a plane wave (\ref{pw}) with $f=*f$ and
$s=-1$ to a plane wave (\ref{pw}) with $f=-*f$.
This basis of free-field states will be used to describe the scattering ---
as in- and out-states in a scattering process.
We shall also adopt the convention that the in- and out-states are
distinguished by the sign of the time component of their 4-momenta $k_\mu$,
namely: $k_0>0$ corresponds to an in-coming particle,
while $k_0<0$ to an out-going one.
Note that the self-duality number $s$ of a plane wave is nothing but the
helicity: $s=1$ means positive helicity for an in-coming photon and negative
helicity for an out-going one (and opposite with $s=-1$).

\medskip

4. Let us discuss now the scattering amplitude
${\cal A}(a_1,a_2,a_3,\dots)$ for an arbitrary number of particles, that is
for in- and out-state given by the plane waves
$a_n=h_n\,e^{ik_n\cdot x} \,,~ n=1,2,3,\dots$, of definite self-dualities
$s_n\,$. We are going to show that
\\
{\it
the tree part of the amplitude,
${\cal A}_{tree}(a_1,a_2,a_3,\dots)$, vanishes unless the sum of self-dualities,
$\sum_ns_n\,$, of all the scattering states is zero.}
\\
This means helicity conservation (in the tree scattering), for $\sum s_n=0$
implies that the sum of helicities in the initial state is the same as in the final state.

Since the connected part of ${\cal A}_{tree}$ is homogeneous with respect to
$a_1,a_2,a_3,\dots$, the above statement amounts to the invariance
with respect to the phase rotations of the plane waves:
\beq \label{aphrot}
{\cal A}_{tree}(e^{is_1\alpha}a_1,e^{is_2\alpha}a_2,e^{is_3\alpha}a_3,\dots)
= {\cal A}_{tree}(a_1,a_2,a_3,\dots)
\eeq
It is this latter form of the conservation law which will be proved below.

\medskip

5. In order to prove this, we recall first the Lehmann-Symanzik-Zimmermann
reduction formula\footnote{
It corresponds to applying amputation on diagrams with one external leg
off shell.}
in the following special form:
\beq \label{LSZ}
{\cal A}_{tree}(a_1,a_2,a_3,\dots) =
\int d^4x\, a_1^\mu(x) \;
\square
A_\mu^{ptb}(x|a_2,a_3,\dots) \,,
\eeq
where $a_1=h_1\,e^{ik_1\cdot x}$ is the plane wave of a chosen, say,
the first scattering particle; while $A_\mu^{ptb}$ is a certain solution
to the classical (non-linear) field equations.
This classical solution, defined in general elsewhere \cite{RS1} under
the name {\it perturbiner} is in fact a generating functional for the
tree-level form-factors of the quantum field $A_\mu(x)$. For the present
needs it is sufficient to mention that $A_\mu^{ptb}$ is a function of the
space-time point $x$ and the quantum numbers $h_n\,,\,k_n$ of the
scattering states, $A^{ptb}=A^{ptb}(x|a_n,\dots)$, and it is uniquely
defined (up to a gauge transformation)
as an expansion in powers of $a_n$'s of the form
\beq \label{expnsn}
A^{ptb}(x) = \sum_n a_n(x) +
{\rm higher~order~terms~in~} a_n{\rm 's} \,,
\eeq
which obeys the classical field equation (\ref{eqm}).

As a classical solution, $A^{ptb}$ is subject to the duality transformations;
for $F^{ptb}=dA^{ptb}$ and $D^{ptb}=D(F^{ptb})$, we have infinitesimally:
\beq
\delta F^{ptb}=*D^{ptb}\,,~~ \delta D^{ptb}=*F^{ptb} \,,
\eeq
or, integrating to a finite rotation\footnote{
Note that we assume that $*^2=1$ and that all the fields are complex, so that
$i=\sqrt{-1}$ in the exponent does not really matter much for the present needs.},
\beq \label{ft}
\begin{array}{lcr}
(F^{ptb}+*D^{ptb}) & \to & e^{i\alpha}(F^{ptb}+*D^{ptb}) \,,\\
(F^{ptb}-*D^{ptb}) & \to & e^{-i\alpha}(F^{ptb}-*D^{ptb}) \,.
\end{array}
\eeq
On the other hand, the field $F^{ptb}(x)$ is uniquely determined by the
quantum numbers of the waves $a_2,a_3,\dots\;$. Therefore, the transformation
(\ref{ft}) should correspond to a transformation of the plane-wave solutions.
The latter can be found by observing the transformation of the first order
term in the expansion (\ref{expnsn}) and turns out to be, of course,
\beq \label{phrot}
a_n \to e^{is_n\alpha} a_n \,,
\eeq
which correspond to the duality rotation of (anti-)self-dual fields
in Maxwell theory.

To summarise, we have the possibility to consider two types of
transformations: the phase rotation of the plane waves, which is applicable
to the functions of $a_1,a_2,a_3,\dots$
(cf.\ eqs.(\ref{phrot}),(\ref{aphrot})) and the duality transformation
(cf.\ eqs.(\ref{it}),(\ref{ft})), which is applicable to the solutions
of the non-linear \BI\ equations. If we denote an infinitesimal phase
rotation of $a_n$'s by $\hat\delta$, we have to prove that
\beq
\hat\delta{\cal A}_{tree}(a_1,a_2,a_3,\dots)=0 \,,
\eeq
while for the perturbiner field $A^{ptb}$ or, rather, for its field
strength, $F^{ptb}$, we have that
\beq
\label{d=d}
\hat\delta F^{ptb} = \delta F^{ptb} \,,
\eeq
where $\delta F^{ptb}=*D^{ptb}$ as earlier.
Thus, to prove our proposition we may apply $\hat\delta$
to both sides of the reduction formula (\ref{LSZ}) and,
then, use the equality (\ref{d=d}) in the right hand side.
Before doing this, we rewrite eq.(\ref{LSZ}) in a more convenient form:
\beq \label{LSZc}
{\cal A}_{tree}(a_1,a_2,a_3,\dots) =
\int *f_1\wedge(F^{ptb}-D^{ptb}) \,.
\eeq
This latter form is obtained from eq.(\ref{LSZ}), in principle, by a formal
integration by parts; one has only to be careful with poles corresponding to
$k_1+k_2+k_3+\dots=0$~\footnote{We have been careful!}.
The rest is easy now:
\beq
\begin{array}{lclcll}
\hat\delta{\cal A}_{tree} & = &
\ds \int \hat\delta*f_1\wedge(F^{ptb}-D^{ptb})         & + &
\ds \int *f_1\wedge(\delta F^{ptb}-\delta D^{ptb}) & =  \\
& = &
\ds \int f_1\wedge(F^{ptb}-D^{ptb}) & + &
\ds \int *f_1\wedge(*D^{ptb}-*F^{ptb}) & =  \\
& = &
\ds \int f_1\wedge(F^{ptb}-D^{ptb}) & + &
\ds \int f_1\wedge(D^{ptb}-F^{ptb}) & =0 \,.
\end{array}
\eeq

\medskip

9. We have just proved the helicity conservation in the \BI\ theory
at the tree level. Note that this means vanishing of sums of certain
diagrams, rather than vanishing of individual diagrams ( helicity is not
preserved by the vertices!). At the one-loop level,
it may be interesting to note that the unitarity implies
vanishing of the  imaginary parts of helicity violating amplitudes.
The latter are then  some {\it rational} functions
of momenta and we would conjecture that it is possible to find
them explicitly, analogously to what was found
in the case of maximally helicity
violating amplitudes in Yang-Mills theory (\cite{who?}).

For the future work we postpone also the question whether helicity
conservation survives quantum corrections in maximally supersymmetric
Born-Infeld theory, as well as applications of our observations to the
string theory.

\bigskip

{\small
\noindent
{\normalsize\bf Acknowledgements}

\medskip
\noindent
This work was supported in part by the Grant INTAS-00-334.
The work of A.R. was also partially supported by the Grant
RFBR-00-02-16530 and the Grant 00-15-96557 for the support
of scientific schools. The work of K.S. was also partially
supported by the Grant CRDF-RP1-2108
and by the Grant 00-15-96562 for the support of scientific schools.
}

\end{document}